\documentclass[runningheads]{llncs}
\usepackage[T1]{fontenc}
\usepackage{multirow}
\usepackage{tabularx}
\usepackage{graphicx}
\usepackage{xcolor}
\usepackage{svg}
\usepackage{bm}
\usepackage{amsmath}
\usepackage{babel}

\usepackage{booktabs}
\usepackage{siunitx}
\usepackage{grffile}
\usepackage{subcaption}
\usepackage{capt-of}
\usepackage{amssymb}
\usepackage{isomath}
\usepackage{paralist}
\usepackage{hyperref}
\usepackage{threeparttable}
\usepackage[capitalise]{cleveref}

\usepackage{subcaption}
\usepackage{xspace}

\begin{document}
\title{Automated Fetal Biometry Assessment with Deep Ensembles using Sparse-Sampling of 2D Intrapartum Ultrasound Images}
\titlerunning{}
%
\author{Jayroop Ramesh*\inst{1}\and
Valentin Bacher*\inst{1}\and
Mark C. Eid*\inst{1,2}\and
Hoda Kalabizadeh*\inst{1}\and
Christian Rupprecht\inst{2}\and 
Ana I.L. Namburete\inst{1}\and 
Pak-Hei Yeung*\inst{1,3}\and
Madeleine K. Wyburd*\inst{1}\and
Nicola K. Dinsdale*\inst{1}}
\authorrunning{J. Ramesh et al.}
%
\institute{Oxford Machine Learning in NeuroImaging Lab, Department of Computer Science, University of Oxford \and
Visual Geometry Group, Department of Engineering Science, University of Oxford \and
College of Computing and Data Science, Nanyang Technological University}
\maketitle              
\begin{abstract}
The International Society of Ultrasound advocates Intrapartum Ultrasound (US) Imaging in Obstetrics and Gynecology (ISUOG) to monitor labour progression through changes in fetal head position. Two reliable ultrasound-derived parameters that are used to predict outcomes of instrumental vaginal delivery are the angle of progression (AoP) and head-symphysis distance (HSD). In this work, as part of the Intrapartum Ultrasounds Grand Challenge (IUGC) 2024, we propose an automated fetal biometry measurement pipeline to reduce intra- and inter-observer variability and improve measurement reliability. 
Our pipeline consists of three key tasks: (i) classification of standard planes (SP) from US videos, (ii) segmentation of fetal head and pubic symphysis from the detected SPs, and (iii) computation of the AoP and HSD from the segmented regions. We perform sparse sampling to mitigate class imbalances and reduce spurious correlations in task (i), and utilize ensemble-based deep learning methods for task (i) and (ii) to enhance generalizability under different US acquisition settings. Finally, to promote robustness in task iii) with respect to the structural fidelity of measurements, we retain the largest connected components and apply ellipse fitting to the segmentations. Our solution achieved ACC: 0.9452, F1: 0.9225, AUC: 0.983, MCC: 0.8361, DSC: 0.918, HD: 19.73, ASD: 5.71, $\Delta_{AoP}$: 8.90 and $\Delta_{HSD}$: 14.35 across an unseen hold-out set of 4 patients and 224 US frames. The results from the proposed automated pipeline can improve the understanding of labour arrest causes and guide the development of clinical risk stratification tools for efficient and effective prenatal care. 

\end{abstract}
\section{Introduction}
Medical ultrasound imaging is routinely used to monitor changes in fetal head position in obstetric exams. 
Among different biometrics, ultrasound parameters angle of progression (AoP) and head-symphysis distance (HSD) are recommended for predicting outcomes of instrumental vaginal delivery \cite{ghi2018isuog}.
In this work, as part of the Intrapartum Ultrasounds Grand Challenge (IUGC) (2024), we propose a pipeline for automated fetal biometry measurement from intrapartum ultrasound videos. The pipeline is formed of three key stages: 1) the detection of standard planes from the input ultrasound videos; 2) the segmentation of both the fetal head and the pubic symphysis regions, and, 3) the calculation of the AoP and HSD from the segmentation masks \cite{bai2022framework,chen2024direction}.

\section{Methods} \label{sec:methods}

\begin{figure}[t]
\centering
\includegraphics[width=0.75\linewidth]{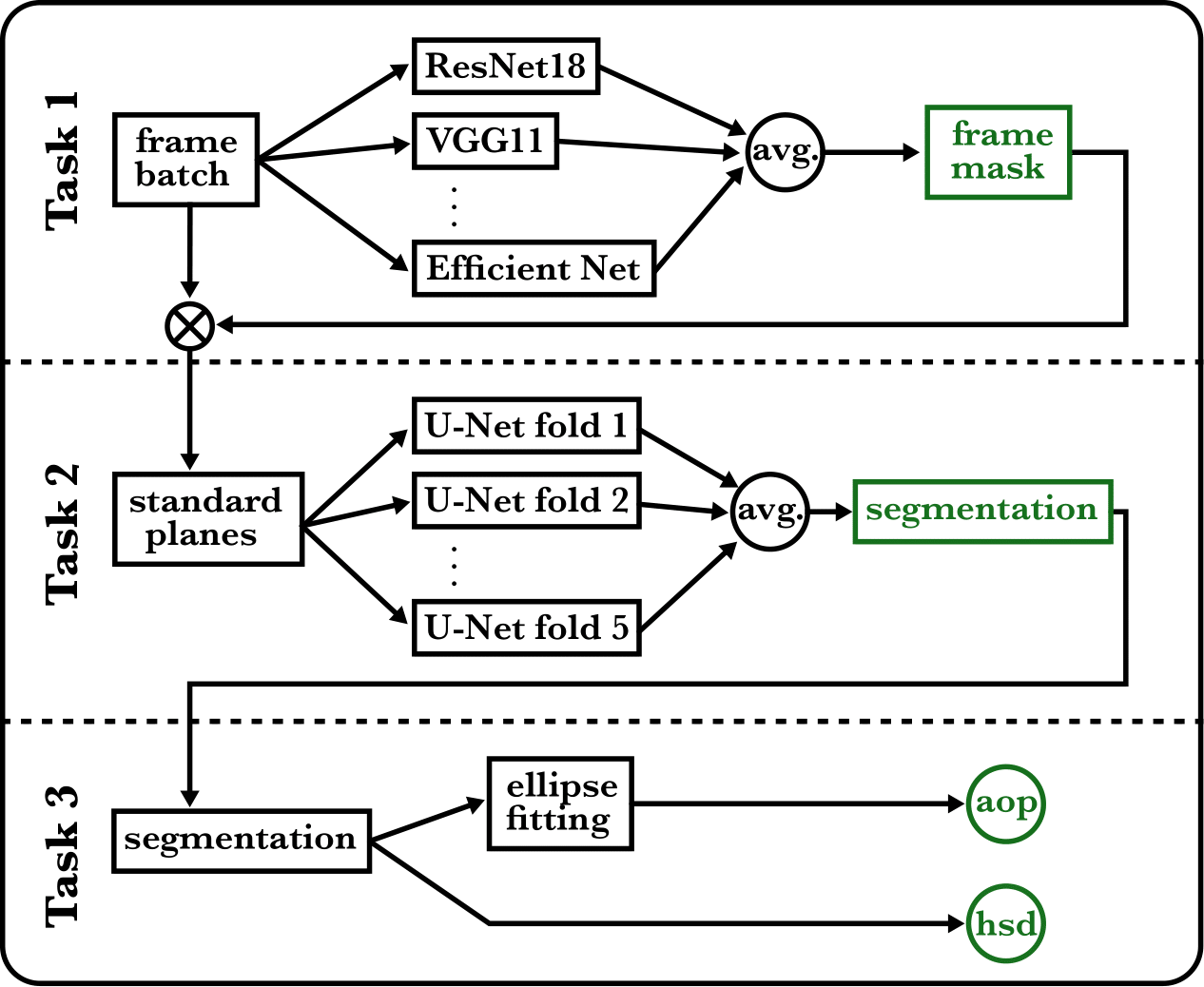}
\caption{Simplified flow chart of a batch of frames being processed by all three tasks. `avg.' is an average operator, ensembling the predictors and $\otimes$ is the masking of the frames to obtain only the standard planes.}\label{fig:methods:flow_chart}
\end{figure}

Figure \ref{fig:methods:flow_chart} shows an overall high-level overview of our automated pipeline.

\subsection{Phase 1: Classification of Standard Planes}

\begin{table}
    \centering
    \begin{tabular}{ll|rl}
    \hline
         Model& Authors & Parameters   \\ \hline
         VGG-11 &\cite{Simonyan2015}&128,780,034 \\
         VGG-16 &\cite{Simonyan2015}&134,277,186 \\
        
        ResNet-18 &\cite{He2015}& 11,177,538  \\
         ResNet-152 &\cite{He2015}& 58,147,906 \\
         WideResNet-50 &\cite{Zagoruyko2017} 
         &66,838,338\\
        DenseNet-196 &\cite{Huang2018}&12,487,810\\ 
        EfficientNet &\cite{Tan2020}&4,010,110\\
        
        EfficientNet-V2 &\cite{Tan2021}&52,860,918\\ 
         ConvNeXT &\cite{Liu2022}&87,568,514\\
         \hline

    \end{tabular}
    \caption{The list of models considered for our classification task. }
    \label{tab:methods:models_classification}
\end{table}

The first phase was the classification of the standard planes from the 2D video streams. We investigated a range of models as shown in Table \ref{tab:methods:models_classification}. All models were either initialized randomly (where explicitly mentioned) or with the ImageNet1K pre-trained weights. We have explored two different training settings, namely (i) training the whole model, and (ii) training just the final classification layer while freezing the rest of the model. It was found that (i) achieved generally superior performance, indicating that more weights required updating to be tailored to this specific downstream task. Therefore, we adopted (i) in our proposed method. To reduce overfitting, a small learning rate of $1 \times 10^{-6}$ was chosen and combined with a learning rate scheduler, halving the learning rate after five epochs of stagnant validation loss.
We summarize the implementations of the models we have investigated as follows:


\begin{enumerate}
    \item \textbf{VGG \cite{Simonyan2015}:} We implemented the architectural variants of VGG-11 (5 convolution-batch norm-pooling layers and 3 fully connected layers), and VGG-16 (8 convolutional layers and 3 fully connected layers) with batch normalization for increased stability and regularization.
    \item \textbf{ResNet \cite{He2015}:} We implemented the architectural variants of ResNet-18 and ResNet-152 (8$\times$ deeper than VGG-16), which are the shallowest and deepest respectively to assess the relative complexity of features important for SP classification.
    \item \textbf{WideResNet \cite{Zagoruyko2017}:} We implemented the architectural variants of WideResNet50, and WideResNet101.
    \item \textbf{DenseNet \cite{Huang2018}:} This approach employs cross-layer connectivity to tackle vanishing gradients, wherein each layer is connected to all layers in the network using a feed-forward configuration. We implemented the architectural variant of DenseNet-196.
    \item \textbf{EfficientNet \cite{Tan2020}:} We implement the baseline model, EfficientNetB0, which consists of depth-wise separable convolutions (MBConv) to reduce computations while retaining representational capacity, as well as squeeze-and-excitation layers which enables the model to focus on emphasizing features while suppressing less relevant ones. We also implement the newer variant EfficientNetV2 \cite{Tan2021}, which utilizes Fused-MBConv that replaces the last layers of MBConv with a simple $3x3$ convolutional layer to speed up training.
    \item \textbf{ConvNeXT \cite{Liu2022}:} ConvNeXt is inspired by elements of transformers but constructed solely using convolutional networks, leveraging depth-wise convolutions, inverted bottlenecks (hidden dimension larger than input dimension), strategically placed larger kernels (non-local self-attention for global receptive fields across layers) for complex modules, activation function selection (replacing ReLU with GERU), and normalization selection (replacing Batch Normalization with Layer Normalization). We implemented the ConvNeXt-Tiny variant.
\end{enumerate}

\subsection{Phase 2: Segmentation of Fetal Head and Pubic Symphysis}

For the video planes classified as standard planes in Phase 1, a model was trained for the segmentation of the Pubic Symphysis (PS) and Fetus Head (FH).

\subsubsection{Models:} 
 For semantic segmentation tasks, capturing both low-level details and high-level information has proven to be highly effective \cite{long2015fully}, and this can be accomplished using a variety of segmentation architectures. The models we have investigated for Phase 2 are shown in \cref{tab:methods:SEGmodels}.

\begin{table}
    \centering
    \begin{tabular}{ll|rl}
    \hline
         Model& Authors & Parameters   \\ \hline
         UNet &\cite{ronneberger2015u}& 486,571  \\
         UNet++ &\cite{zhou2019unet++}&(15,970,739, 33,858,931) \\
         DeepLabV3+ &\cite{chen2018encoder}& (4,907,983, 13,353,981)  \\
         MA-Net &\cite{fan2020ma}& 23,927,027   \\
         \hline

    \end{tabular}
    \caption{The list of models considered for our segmentation task. }
    \label{tab:methods:SEGmodels}
\end{table}

\begin{enumerate}
    \item \textbf{UNET \cite{ronneberger2015u}}: The selected U-Net architecture consisted of four encoding and decoding layers. Each layer is comprised of a convolutional layer, a ReLU activation function, and a batch normalisation layer. The depth, determined by the number of convolutional filters, doubled with each successive layer, starting from 4.
    \item \textbf{UNet++ \cite{zhou2019unet++}:} The architecture consisted of 5 encoding and decoding layers and used a pre-trained ResNet-18 or a VGG-11 backbone for the encoder across two implementations. The decoder also includes batch normalisation and uses progressively decreasing channels for each layer [256, 128, 64, 32, 16].
    \item \textbf{DeepLabV3+ \cite{zhou2019unet++}:} The architecture consisted of an encoder with depth 5, an output stride of 16, and used either a pre-trained EfficientNetB0 or a DPN-68b \cite{chen2017dual} backbone across two implementations. The decoder channels are fixed at 256, and an upsampling factor is applied at the final step. It also leverages atrous/dilated convolution with 3 dilation rates to capture context at differing scales: [12, 24, 36].
    \item \textbf{MA-Net \cite{fan2020ma}:} The architecture consisted of a pre-trained MIT-B1\cite{yu2023mix} encoder with a depth of 5, batch normalization and progressive attention blocks with fixed 64 channels per block. The decoder channels progressively decrease in the order [256, 128, 64, 32, 16]. 
\end{enumerate}

\subsubsection{Postprocessing:}
Postprocessing was applied to the output of the segmentation model to refine the masks and remove any incorrect segmentation outside of the two main regions of interest. This was achieved by retaining only the largest connected components for each class label and discarding any smaller disconnected ones.
    
\subsection{Phase 3: Measurement of Ultrasound Parameters}

The final stage of the pipeline is the measurement of key parameters from the predicted segmentation masks from Phase 2. We perform the following steps to find the necessary landmarks on the PS area and FH and get accurate AoP and Head-Symphysis Distance HSD measurements. The following steps are applied to both segmentations, FH and PS separately, using OpenCV.

\begin{enumerate}
    \item As the predicted segmentation mask may contain some holes, we first apply \textit{morphological closing} with a $10 \times 10$ pixel elliptical kernel. Closing is a two-step process. The segmentation is first dilated (to close up any holes by expanding the entire segmentation) and then eroded (to return the segmentation mask roughly to its original size apart from the closed holes).
    
    \item We then apply \textit{Canny edge-detection}, to detect and smooth the edges of the two segmentation masks. We use hysteresis threshold values of minVal = 2 and maxVal = 5. A pixel on an edge under consideration with an intensity gradient $> \! \text{maxVal}$ is straight away classified as being part of an edge (i.e. a ``sure edge''), whilst one with an intensity gradient $< \! \text{minVal}$ is classified as not being part of an edge (i.e. a ``non-edge''). In between, if the edge pixel is directly connected to a ``sure-edge'' pixel, then it is also classed as an edge pixel, otherwise a ``non-edge'' pixel.
    
    \item If multiple edges (\emph{large interior holes not closed in step 1} are detected, the longest edge will be selected as the boundary for the PS/FH. 
    An Approximate Mean Squared (AMS) ellipse is then fitted to that boundary, and filled in. As both the PS and FH are roughly elliptical, we believe it is more accurate to enforce that their segmentation must also be elliptical.

    \begin{figure}[t]
        \centering
        \includegraphics[width=0.6\linewidth]{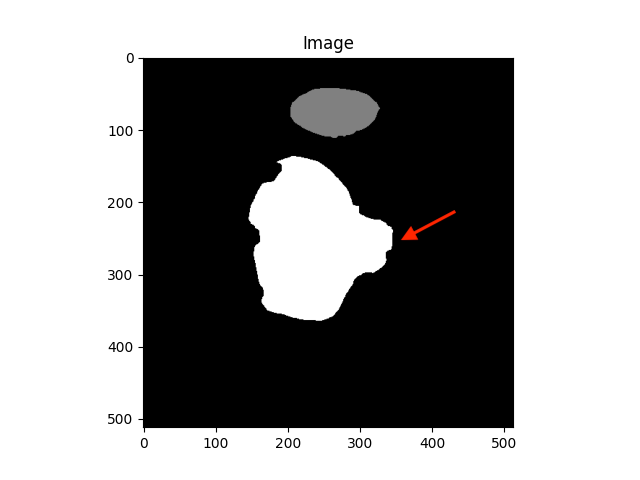}
        \caption{An example of a ``protrusion'' (shown by the red arrow) in some of the predicted segmentation.}
        \label{fig:protrusion example}
    \end{figure}
    
    \begin{figure}[t]
        \centering
        \includegraphics[width=0.6\linewidth]{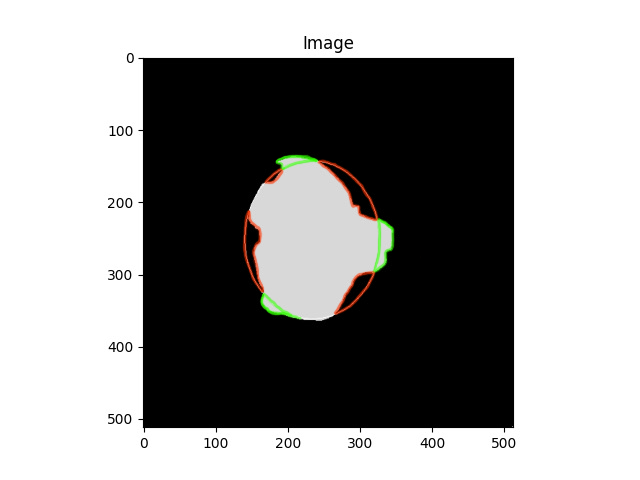}
        \caption{For this segmentation, $E \cup \overline{S}$ is the 4 areas enclosed by the \textbf{red} highlights. $\overline{E} \cup S$ is the 3 areas enclosed by the \textbf{green} highlights}
        \label{fig:E and S diagram}
    \end{figure}
    
    \item The segmentation masks predicted in Phase 2 may contain large protrusions (\cref{fig:protrusion example}), which can greatly distort the fitting of the ellipse. To address this, We apply the following condition: the number of pixels enclosed by the fitted ellipse but not part of the segmentation (red region in \cref{fig:E and S diagram}) must be less than the number of segmentation mask pixels located outside the fitted ellipse (green region in \cref{fig:E and S diagram}). 
    This ensures that the fitted ellipse is not unnecessarily large or skewed to accommodate anomalous segmentation protrusions. This condition can be summarized as:
    \begin{equation}
        \begin{aligned}
            &\text{If } \frac{E \cup \overline{S}}{\overline{E} \cup S} < 1 \text{:} \\
            &\hspace{1em} \text{Protrusion not/no longer detected} \nonumber \\
            &\text{Else:} \nonumber \\
            &\hspace{1em} \text{Protrusion is/is still detected} \nonumber
        \end{aligned}
    \end{equation}
    Where $E$ is the area enclosed by the FH/PS fitted AMS ellipse, and $S$ is the area enclosed by the hole-closed predicted FH/PS segmentation mask.\\
    
    When a protrusion is detected, we iteratively prune the pixels from the segmentation mask that extend beyond a user-specified threshold outside the fitted ellipse. A new ellipse is then fitted to the pruned segmentation, and the presence of protrusions is reassessed. If protrusions persist, the pruning process is repeated until 15 pruning iterations have been performed, or no further protrusions are detected, whichever occurs first. This operation is summarized in \cref{fig:iterative pruning} and as follows:
    \begin{equation}
        \begin{aligned}
            &\text{while } \frac{E \cup \overline{S}}{\overline{E} \cup S} > 1 \text{ and } count\leq15 \text{:} \\
            &\hspace{1em} \text{(Continue to) Prune segmentation} \nonumber \\ &\hspace{12pt} count\mathrel{+}=1 \nonumber \\
        \end{aligned}
    \end{equation}
    where $E$ is the area enclosed by the FH/PS fitted AMS ellipse, and $S$ is the area enclosed by the hole-closed predicted FH/PS segmentation mask.
    \begin{figure}[]
        \centering
        \includegraphics[width=1\linewidth]{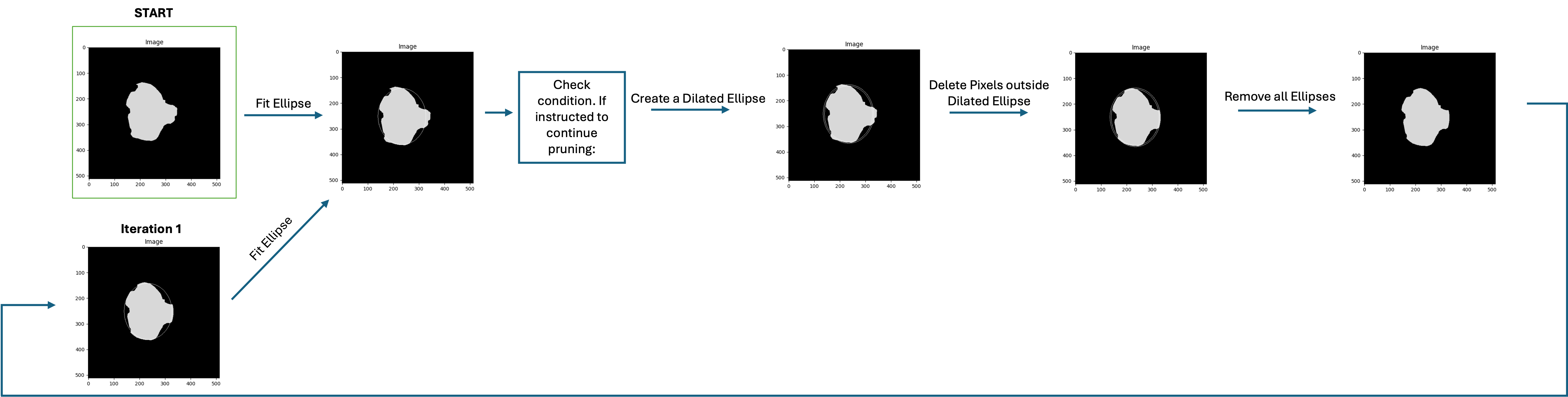}
        \caption{Diagram of the iterative pruning process as described in Step 4.}
        \label{fig:iterative pruning}
    \end{figure}
    \\
    \item In some cases we found that the approximation or assumption that the PS and FH are elliptical resulted in less accurate biometric measurements compared to the original, hole-closed predicted segmentation. Therefore, we implemented a decision rule to determine whether to use the fitted ellipse or the original predicted segmentation, for both the PS and FH independently:
    \begin{equation}
        \begin{aligned}
            &\text{If } \frac{E \cup \overline{S}}{S} < 20\% \text{:} \\
            &\hspace{1em} \text{Use fitted AMS ellipse} \nonumber \\
            &\text{Else:} \nonumber \\
            &\hspace{1em} \text{Use hole-closed predicted segmentation} \nonumber
        \end{aligned}
    \end{equation}
    where $E$ is the area enclosed by the FH/PS fitted AMS ellipse, $S$ is the area enclosed by the hole-closed predicted FH/PS segmentation mask, and $(E \cup \overline{S})$ is the red region in \cref{fig:E and S diagram}.
    \vspace{0.5\baselineskip} 

    \item As decided upon in Step 5, the most suitable segmentation masks for the FH and PS are used to derive the AoP measurement. For calculating HSD, We found it more accurate to \textbf{avoid} assuming elliptical shapes for the PS and FH. Instead, we used only the hole-closed predicated segmentations (\emph{i.e.} the output of Step 1) to identify the two relevant landmarks and compute the HSD. 
\end{enumerate}

\section{Data}
\subsubsection{Phase 1: Classification of Standard Planes}

To train the final model, we employed sparse sampling, using only a fraction of frames from each video. Training with the entire video led to rapid overfitting, impacting the models' generalization to validation and testing data. One possible explanation for this is that in the training data, entire videos and thus subjects were labelled positive or negative. Consequently, the model could ``cheat'' by memorizing the subjects rather than learning meaningful patterns for classification \cite{Mitchell2019}.
To reduce the correlation between label and subject, we only sampled a fixed number of frames from each positive or negative video,  significantly improving training performance. The final dataset consisted of 5 frames of each positive-labelled video and 8 frames of each negative-labelled frame to have equally balanced numbers of positive and negative frames. As testing will be performed on a separate set of videos, we also used a part of the provided validation data to train our model. 

All classification models were trained using the AdamW optimiser for 100 epochs with an initial learning rate of $10^{-4}$, following which it decreased by a factor of 0.1 if the validation performance plateaued for 5 epochs. Early stopping was employed if no improvement in validation performance was observed for 15 epochs. The loss function was formed of a combination of Binary Cross Entropy (BCE) with optimizer parameters $\beta_{0} = 0.9, \beta_{1} = 0.999$ chosen empirically.

\begin{figure}
    \centering
    \includegraphics[width=1\linewidth]{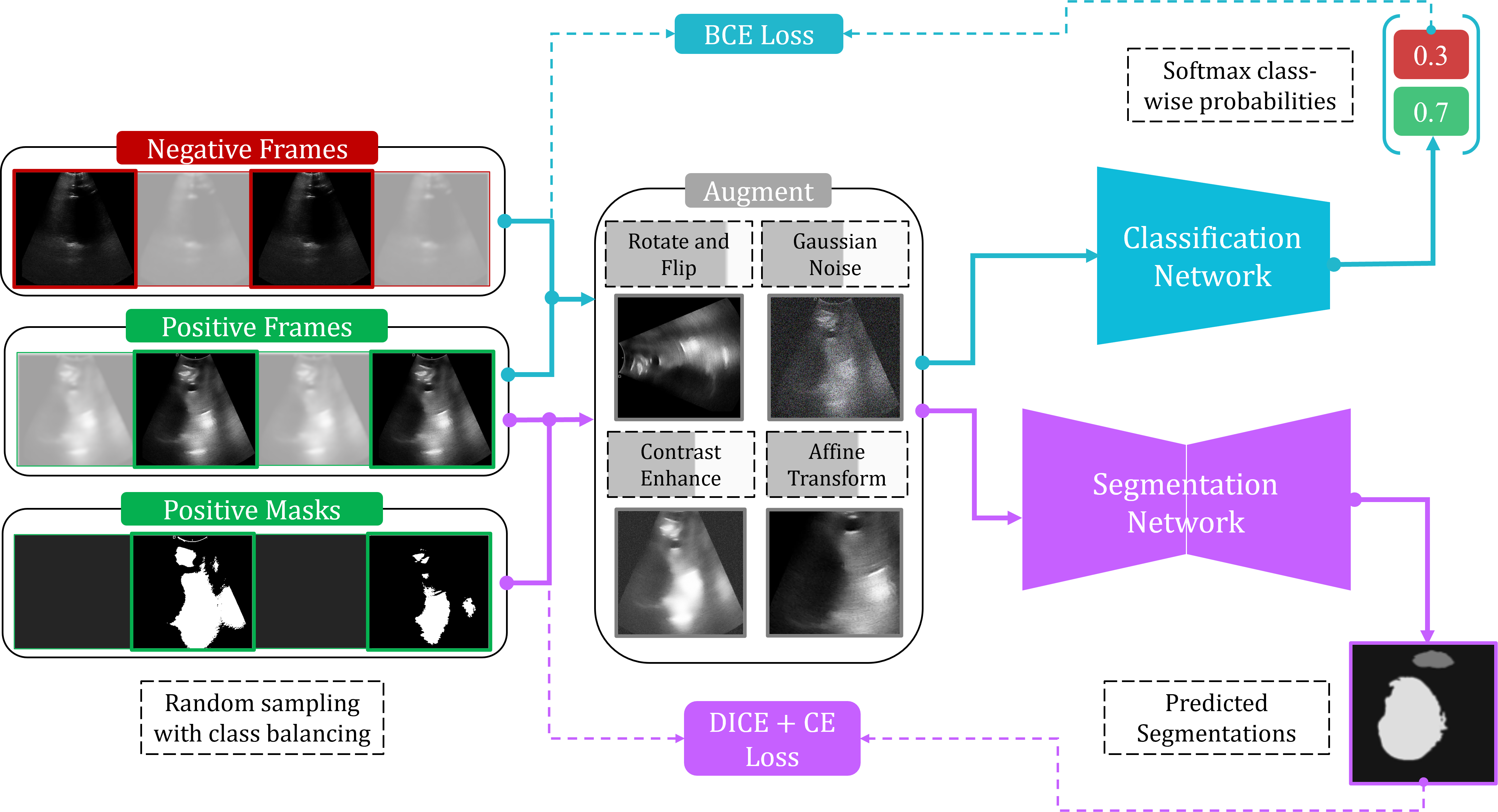}
    \caption{Overview of the training process, from spare-sampling to augmentation, followed by classification/segmentation.}
    \label{fig:training_pipeline}
\end{figure}

\subsubsection{Phase 2: Segmentation of Fetal Head and Pubic Symphysis}

The training data was inherently arranged such that only a few SP frames from a positively labelled video had corresponding segmentation masks. On average, each patient-specific video contained between 5 and 7 segmentation masks, leading to a training set of 1,987 image-mask pairs indicating SP. As mentioned previously, we also used a part (90\%) of the provided validation data to train our model, where each video had only 1 segmentation mask corresponding to the highest quality SP frame.

All segmentation models were trained using the Adam optimiser for 180 epochs with an initial learning rate of $10^{-3}$ for 100 epochs, following which it decreased linearly for the remaining 80. The loss function was formed of a combination of Cross Entropy (CE) and Dice loss as per (\ref{eq:loss_Seg}) with $\lambda_{1} = \lambda_{2} = 0.5$, with the value chosen empirically. 

\begin{equation}
    L_{seg} = \lambda_{1} L_{CE}
    + \lambda_{2}L_{Dice}
    \label{eq:loss_Seg}
\end{equation}

\textbf{Testing:} We used 4 videos of randomly selected individuals in the validation dataset (10\%) for our validation which was mostly used for the learning rate scheduler to prevent overfitting. The rest of the validation dataset was used for training as well as it was provided with per-frame labels. This process is depicted in \cref{fig:training_pipeline}.

\subsection{Augmentation}
Training data augmentation helps facilitate the model's ability to generalise to unseen training data. When training our proposed models, random data augmentations were applied, which are summarized in \cref{tab:methods:transforms}. Intensity-based augmentations alter the intensity values of the ultrasound images. Gaussian noise, contrast adjustment and gamma enhancement fall under this category. Additive Gaussian noise adds randomly sampled values to the original image values. The added Gaussian noise typically has zero mean and a small standard deviation when applied to images with grey values between 0 and 1. Gamma enhancement adjusts the dynamic range of an image by feeding the image values through an exponential function, stretching or compressing different parts of the spectrum. Contrast augmentations mainly clip the grey value range or expand it linearly. Affine augmentations move and alter the pixel space relative to the image frame. Augmentation ranges (\cref{tab:methods:transforms}) were carefully selected such that they are typical for ultrasound images to avoid unnecessarily complicating the task by introducing unrealistically large variations, while still ensuring sufficient generalization to unseen data.

For classification, all augmentations were applied. For segmentation, the affine translation and scaling were not applied and the corresponding masks were also adjusted for consistency when flips/rotations were used.

While augmentations were only applied to training data, both training and testing data were preprocessed by linearly mapping the image intensity per channel from the range of 0-255 to 0-1, to allow for the use of ImageNet1K-pre-trained models.

\begin{table}[t]
    \centering
    \begin{tabular}{l|ll}
    \hline
         Augmentation&  Range& Probability\\ \hline
         Horizontal flip& - & (0.5-0.8)\\
         Additive Gaussian Noise& $\mu = 0, \, \sigma \in [1.18 \times 10^{-2}, 5.88 \times 10^{-2}]$ & 0.5\\
         Gamma Enhancement& $\text{image}^{0.4 \text{ to } 1}$ & 0.5\\
         Contrast& [0.8,1.2] & 0.5\\
         Affine&  & 0.6\\
         \hspace{6pt} Translation&  $[-0.1, 0.1] \times \text{W/ H}$ & \\
         \hspace{6pt} Rotation& $\pm 20 \deg$ & \\
         \hspace{6pt} Scaling& [1, 1.3] & \\\hline
    \end{tabular}
    \caption{This table lists the augmentation techniques' ranges and their probabilities. For these values, the images are assumed to have grey values within the range of 0 and 1.}
    \label{tab:methods:transforms}
\end{table}

\section{Results}

For Phase 1, we use Accuracy (ACC), the harmonic mean of precision and recall, \emph{i.e.}, the F1 score, Area under the Receiver Operating Characteristic curve (AUC), and Matthews Correlation Coefficient (MCC) to evaluate classification. For Phase 2, we used the Dice similarity coefficient (DSC), Average Surface Distance (ASD) and Hausdorff distance (HD) to evaluate the segmentation. For Phase 3, the AoP and HSD were used to evaluate fetal biometry measurement. All pre-trained models were implemented using timm (classification) \cite{rw2019timm} and SMP (segmentation) \cite{Iakubovskii:2019} libraries.

We assess the performance of different individual models for classification in \cref{tab:classification_indx} and segmentation in \cref{tab:seg-results}. Given the input image $x$ of size [512,512,3] and output of negative and class softmax probabilities $y$, we ensemble them as follows in Equation \ref{Eq:ens}: 

\begin{equation}
p(y \mid \mathbf{x})=M^{-1} \sum_{m=1}^M p_{\theta_m}\left(y \mid \mathbf{x}, \theta_m\right) 
\label{Eq:ens}
\end{equation}

Where this uniformly weighted mixture model corresponds to averaging the predicted
probabilities (in the case of classification) or predicted mask pixel-wise values (in the case of segmentation) of the $M$ individual models. It is worth mentioning that majority voting was also implemented, but averaging produced superior results across both tasks. 

\subsection{Phase 1: Classification of Standard Planes}

The ensemble model (CLF-5) achieved the highest average \textbf{ACC: 0.9452, F1: 0.9225, AUC: 0.983 and MCC: 0.8361} with a total trainable parameter count of \textbf{244,024,006}. While in this limited testing set, EfficientNetB0 fares slightly better ($<1\%$, ensembles tend to generalize better due to a combination of each model's inductive biases \cite{jr-lakshminarayanan2017simple,kendall2017uncertainties}). Furthermore, diverse ensembles can complement missing gaps in knowledge information, and if fine-tuned on a new task, lead to a less redundant, multiple low-loss regions \cite{jr-fort2019deep,valdenegro2022deeper}.

\begin{table}
    \centering
     \caption{This table contains evaluation metrics for classification on the testing data of the individual models. The best scores are in \textbf{bold}, and the second best scores are \underline{underlined}. Upward arrows ($\uparrow$) indicate higher value is better and downward arrows ($\downarrow$) indicate lower value is better for metrics. }
    \begin{threeparttable}
    \begin{tabular}{l|llll}
\hline
         Model  & Accuracy $(\uparrow)$ & F1 $(\uparrow)$ & AUC $(\uparrow)$ & MCC $(\uparrow)$  \\ \hline
   
         ResNet-18*  & 0.927 & \underline{0.894} & 0.992 & \underline{0.800}  \\
         ResNet-18 & 0.918 & 0.873 & \underline{0.995} & 0.783  \\
         VGG-11* & 0.829 & 0.797 & \textbf{0.997} & 0.673  \\
         VGG-11  & \underline{0.928} & 0.881 & 0.981 & 0.766  \\
         VGG-16  & 0.923 & 0.877 & 0.981 & 0.758  \\
         WideResNet-50  & 0.847 & 0.803 & 0.982 & 0.664  \\ 
         DenseNet-196  & 0.900 & 0.842 & 0.974 & 0.714  \\
         EfficientNetB0 &\textbf{0.951} & \textbf{0.931} & 0.986 & \textbf{0.848}  \\
         EfficientNet-V2  & 0.899 & 0.831 & 0.947 & 0.631  \\
         ConvNeXt  & 0.923 & 0.881 & 0.984 & 0.795  \\\hline
    \end{tabular}
    \begin{tablenotes}
        \footnotesize
        \item[*] The asterisk indicates models trained from scratch using random initialisation. All other models were trained using full fine-tuning. Worse-performing variants are not reported. 
    \end{tablenotes}
    \end{threeparttable}
    \label{tab:classification_indx}
\end{table}

\begin{table}
    \centering
       \caption{This table contains evaluation metrics on the testing data of the ensemble classification models. The best scores are in \textbf{bold}, and the second best scores are \underline{underlined}. Upward arrows ($\uparrow$) indicate higher value is better and downward arrows ($\downarrow$) indicate lower value is better for metrics. "CLF$-M$" indicates an ensemble of $M$ models.}
    \begin{tabular}{lc|llll}
    \hline
         Ensemble & Models & Accuracy $(\uparrow)$ &F1 $(\uparrow)$  & AUC $(\uparrow)$ & MCC $(\uparrow)$ \\ \hline
         CLF-2 & EfficientNetB0 + ConvNeXt & 0.927  & 0.887 & 0.988 &  0.788\\
         CLF-3 & CLF-2 + ResNet-18* &  0.937& 0.918 & 0.986 & 0.801 \\
         CLF-4 & CLF-3 + VGG-11 &  0.925& 0.883 & \textbf{0.989} & 0.798   \\
         CLF-5 & CLF-4 + DenseNet-196 & \textbf{0.945}&  \textbf{0.922}& \underline{0.983}&\textbf{0.836 }  \\
         CLF-7 & CLF-5 + VGG-16 + EfficientNet-V2  & 0.923 & 0.891 & 0.988 & 0.770   \\ \hline
    \end{tabular}
 
    \label{tab:classification_ens}
\end{table}

\begin{figure}
    \centering
    \includegraphics[width=0.6\linewidth]{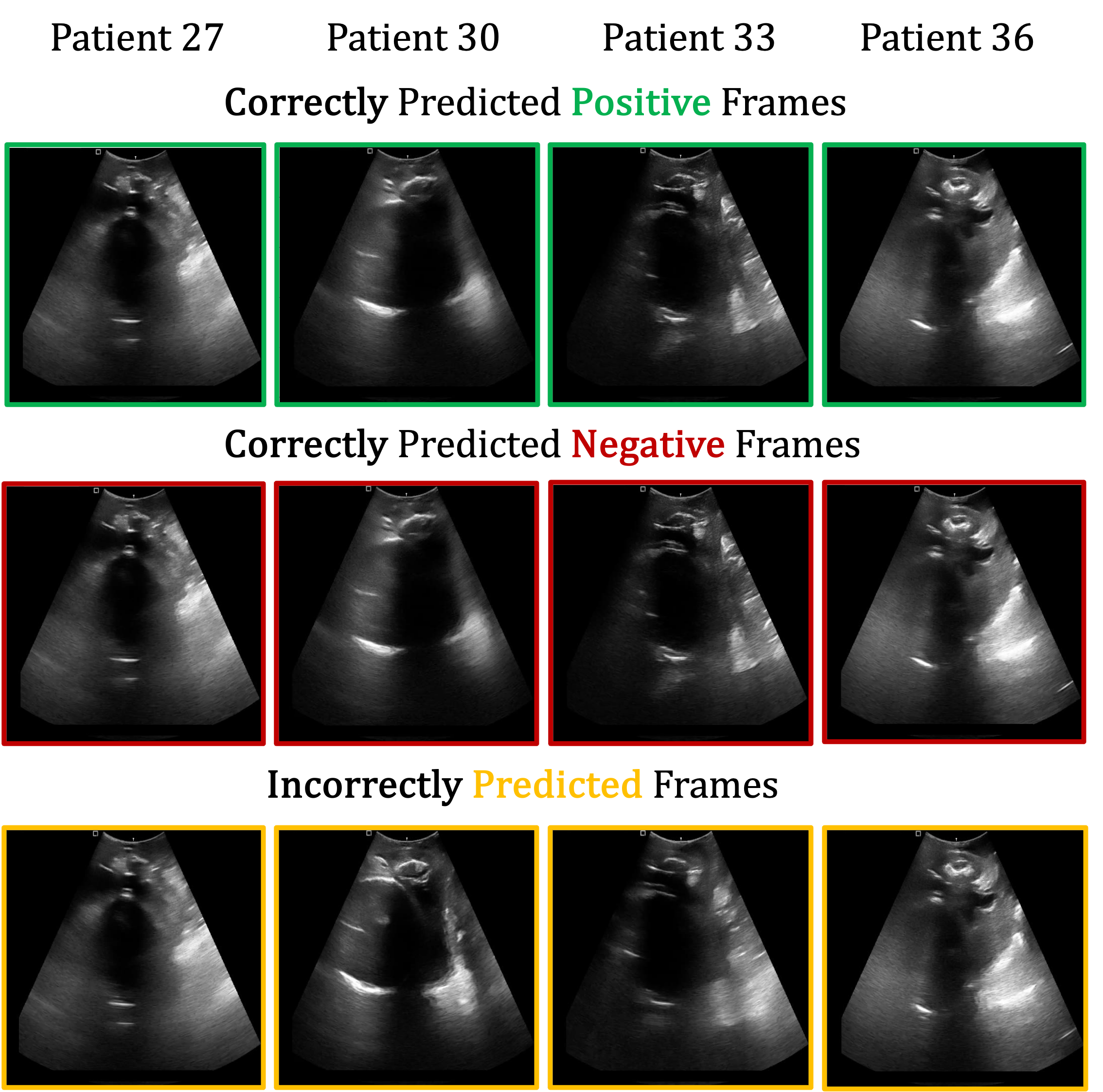}
    \caption{Predicted SPs from Phase 1, True Positives on Row 1, True Negatives on Row 2 and a combination of False Positives and Negatives on Row 3.}
    \label{fig:classification_images}
\end{figure}

\subsection{Phase 2: Segmentation of Fetal Head and Pubic Symphysis}
\begin{figure}
    \centering
    \includegraphics[width=0.9\linewidth]{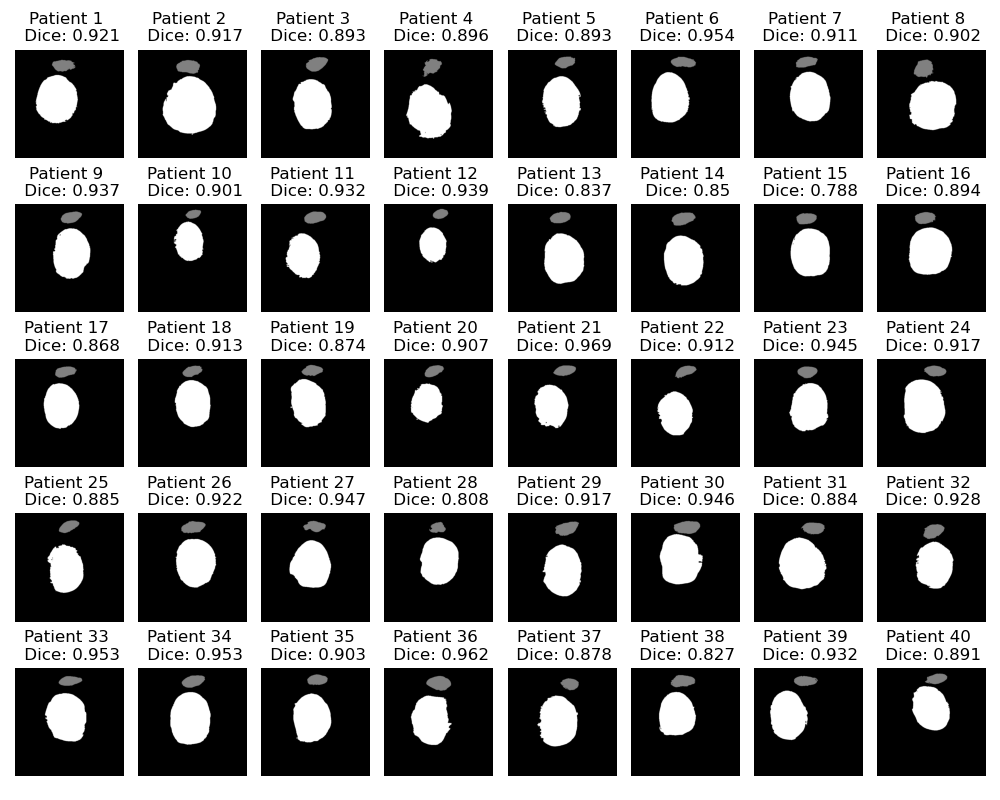}
    \caption{Predicted segmentation masks and corresponding Dice scores.}
    \label{fig:segmentation_masks}
\end{figure}

The segmentation masks predicted by the best-performing model (SEG-3) have been visualised in Figure \ref{fig:segmentation_masks}. This model achieved the highest average \textbf{Dice score of 0.919, and the lowest ASD and HD of 5.712 and 19.735 respectively} with a total trainable parameter count of \textbf{62,693,941.}

\begin{table}
\centering
\caption{This table contains evaluation metrics for segmentation on the testing data of the individual models. The best scores are in \textbf{bold}, and the second best scores are \underline{underlined}. Upward arrows ($\uparrow$) indicate higher value is better and downward arrows ($\downarrow$) indicate lower value is better for metrics.}
\label{tab:seg-results}
\begin{tabular}{l|lllll}
\hline
Model & DSC $(\uparrow)$ & ASD $(\downarrow)$ & HD $(\downarrow)$ & $\Delta_{AoP}$ $(\downarrow)$ & $\Delta_{HSD}$ $(\downarrow)$ \\ \hline
UNet & 0.883 & 8.494 & 29.831 & 8.801 & \underline{8.077} \\
Unet++ (ResNet-18) & 0.881 & 8.179 & 25.082 & \textbf{8.036} & 20.220 \\
Unet++ (VGG-11) & \underline{0.899} & \underline{6.467} & \underline{24.853} & 12.036 & 11.480 \\
DeepLabv3 (EfficientNetB0) & 0.862 & 9.462 & 25.691 & 11.724 & 11.960 \\
DeepLabv3 (DPN-68B) & 0.868 & 9.818 & 30.220 & 10.952 & \textbf{5.908} \\
MA-Net (MIT-B1) & \textbf{0.914} & \textbf{5.936} & \textbf{21.618} & \underline{8.442} & 17.815 \\ \hline
\end{tabular}
\end{table}

\begin{table}
    \centering
    \caption{This table compares the errors on the two biometric measurements AoP and HDS. The errors are calculated between the raw predictions (Pred) and using ellipse fitting (Ellips) relative to the predictions on the ground truth segmentations. The smaller value between each error of each measurement is highlighted in \textbf{bold}}.
    \label{tab:eval_elips_fitting}
    \begin{tabular}{c|cc|cc}
    \hline
         Models & $\Delta^{\text{Pred}}_{\text{AoP}}$ & $\Delta^{\text{Ellips}}_{\text{AoP}}$ & $\Delta^{\text{Pred}}_{\text{HSD}}$ & $\Delta^{\text{Ellips}}_{\text{HSD}}$\\ \hline
         U-Net  & \textbf{10.571} & 11.659 & 13.331 & \textbf{11.516}\\
         SEG-3  & \textbf{10.571} & 12.653 & 13.331 & \textbf{12.053}\\ \hline
    \end{tabular}
\end{table}

\subsection{Phase 3: Measurement of Ultrasound Parameters} \label{sec:Phase 3 Method}
We found that SEG-3 produced the best segmentations and measurements, so this is what was used when calculating the measurements. Using the predicted segmentations and our methodology detailed in Section \ref{sec:Phase 3 Method}, AoP and HSD were calculated on 4 held-out test patients. The difference between the ground truth and predicted AoP and HSD was calculated for each of these 4 patients. The results of the average errors are presented in \cref{tab:eval_elips_fitting}. As these measurements heavily depend on the segmentation results, the choice of only using ellipse fitting for HSD is optimal for our segmentation but not a general result. \textbf{The average $\Delta_{AoP}$ was 8.906\textdegree, and the average $\Delta_{HSD}$ was 14.356 pixels}.

\begin{table}
    \centering
       \caption{This table contains evaluation metrics on the testing data of the ensemble segmentation models. The best scores are in \textbf{bold}, and the second best scores are \underline{underlined}. Upward arrows ($\uparrow$) indicate higher value is better and downward arrows ($\downarrow$) indicate lower value is better for metrics. "SEG$-M$" indicates an ensemble of $M$ models.}
    \begin{tabular}{lc|lll}
    \hline
         Ensemble & Models & DSC $(\uparrow)$ & ASD $(\downarrow)$ & HD $(\downarrow)$ \\ \hline
         SEG-2 & DeepLabV3+(EfficientNetB0) + MA-Net(MIT-B1) &  0.912  & \textbf{5.442} & \underline{21.659} \\
         SEG-3 & SEG-2 + UNet++(VGG-11) & \textbf{0.918}& \underline{5.712}&\textbf{19.734 } \\
         SEG-4 & SEG-3 + UNet & 0.909 &  6.350& 22.326  \\
         SEG-5 & SEG-4 + DeepLab(DPN-68B) &0.906 &6.658 & 22.175 \\
         SEG-6 & SEG-5 + UNet++(ResNet-18)  &0.903 &7.059 & 23.013 \\
         \hline
    \end{tabular}
    \label{tab:results:classification}
\end{table}

\section{Conclusion}
\label{sec:conc}
We have developed a pipeline for the calculation of key US parameters for measuring outcomes of instrumental vaginal delivery, starting from intrapartum ultrasound videos. The three-stage pipeline identifies the standard planes, segments the fetal head and pubic symphysis regions, and then calculates the key ultrasound parameters. Throughout, care has been taken to develop models which will generalise well to the range of expected US videos, and our results show the potential of using automated methods for supporting prenatal care. 

\section*{Acknowledgements}
\label{sec:ack}
The authors acknowledge the generous support of the EPSRC Centre for Doctoral Training in Autonomous Intelligent Machines \& Systems (EP/S024050/1), Amazon Web Services, Oxford Department of Computer Science Scholarship, the Bill \& Melinda Gates Foundation, the Presidential Postdoctoral Fellowship from the Nanyang Technological University, and the EPSRC Doctoral Prize Scheme.

\bibliographystyle{splncs04}
\bibliography{references}

\begin{thebibliography}{10}
\providecommand{\url}[1]{\texttt{#1}}
\providecommand{\urlprefix}{URL }
\providecommand{\doi}[1]{https://doi.org/#1}

\bibitem{bai2022framework}
Bai, J., Sun, Z., Yu, S., Lu, Y., Long, S., Wang, H., Qiu, R., Ou, Z., Zhou, M., Zhi, D., et~al.: A framework for computing angle of progression from transperineal ultrasound images for evaluating fetal head descent using a novel double branch network. Frontiers in physiology  \textbf{13},  940150 (2022)

\bibitem{chen2018encoder}
Chen, L.C., Zhu, Y., Papandreou, G., Schroff, F., Adam, H.: Encoder-decoder with atrous separable convolution for semantic image segmentation. In: Proceedings of the European conference on computer vision (ECCV). pp. 801--818 (2018)

\bibitem{chen2017dual}
Chen, Y., Li, J., Xiao, H., Jin, X., Yan, S., Feng, J.: Dual path networks. Advances in neural information processing systems  \textbf{30} (2017)

\bibitem{chen2024direction}
Chen, Z., Ou, Z., Lu, Y., Bai, J.: Direction-guided and multi-scale feature screening for fetal head--pubic symphysis segmentation and angle of progression calculation. Expert Systems with Applications  \textbf{245},  123096 (2024)

\bibitem{Mitchell2019}
Dawson, M., Zisserman, A., Nell{\aa}ker, C.: From same photo: Cheating on visual kinship challenges. In: Jawahar, C.V., Li, H., Mori, G., Schindler, K. (eds.) Computer Vision -- ACCV 2018. pp. 654--668. Springer International Publishing, Cham (2019)

\bibitem{fan2020ma}
Fan, T., Wang, G., Li, Y., Wang, H.: Ma-net: A multi-scale attention network for liver and tumor segmentation. IEEE Access  \textbf{8},  179656--179665 (2020)

\bibitem{jr-fort2019deep}
Fort, S., Hu, H., Lakshminarayanan, B.: Deep ensembles: A loss landscape perspective. arXiv preprint arXiv:1912.02757  (2019)

\bibitem{ghi2018isuog}
Ghi, T., Eggeb{\o}, T., Lees, C., Kalache, K., Rozenberg, P., Youssef, A., Salomon, L., Tutschek, B.: Isuog practice guidelines: intrapartum ultrasound. Ultrasound in Obstetrics \& Gynecology  \textbf{52}(1),  128--139 (2018)

\bibitem{He2015}
He, K., Zhang, X., Ren, S., Sun, J.: Deep residual learning for image recognition (2015), \url{https://arxiv.org/abs/1512.03385}

\bibitem{Huang2018}
Huang, G., Liu, Z., van~der Maaten, L., Weinberger, K.Q.: Densely connected convolutional networks (2018), \url{https://arxiv.org/abs/1608.06993}

\bibitem{Iakubovskii:2019}
Iakubovskii, P.: Segmentation models pytorch. \url{https://github.com/qubvel/segmentation_models.pytorch} (2019)

\bibitem{kendall2017uncertainties}
Kendall, A., Gal, Y.: What uncertainties do we need in bayesian deep learning for computer vision? Advances in neural information processing systems  \textbf{30} (2017)

\bibitem{jr-lakshminarayanan2017simple}
Lakshminarayanan, B., Pritzel, A., Blundell, C.: Simple and scalable predictive uncertainty estimation using deep ensembles. Advances in neural information processing systems  \textbf{30} (2017)

\bibitem{Liu2022}
Liu, Z., Mao, H., Wu, C.Y., Feichtenhofer, C., Darrell, T., Xie, S.: A convnet for the 2020s (2022), \url{https://arxiv.org/abs/2201.03545}

\bibitem{long2015fully}
Long, J., Shelhamer, E., Darrell, T.: Fully convolutional networks for semantic segmentation. In: Proceedings of the IEEE conference on computer vision and pattern recognition. pp. 3431--3440 (2015)

\bibitem{ronneberger2015u}
Ronneberger, O., Fischer, P., Brox, T.: U-net: Convolutional networks for biomedical image segmentation. In: Medical image computing and computer-assisted intervention--MICCAI 2015: 18th international conference, Munich, Germany, October 5-9, 2015, proceedings, part III 18. pp. 234--241. Springer (2015)

\bibitem{Simonyan2015}
Simonyan, K., Zisserman, A.: Very deep convolutional networks for large-scale image recognition (2015), \url{https://arxiv.org/abs/1409.1556}

\bibitem{Tan2020}
Tan, M., Le, Q.V.: Efficientnet: Rethinking model scaling for convolutional neural networks (2020), \url{https://arxiv.org/abs/1905.11946}

\bibitem{Tan2021}
Tan, M., Le, Q.V.: Efficientnetv2: Smaller models and faster training (2021), \url{https://arxiv.org/abs/2104.00298}

\bibitem{valdenegro2022deeper}
Valdenegro-Toro, M., Mori, D.S.: A deeper look into aleatoric and epistemic uncertainty disentanglement. In: 2022 IEEE/CVF Conference on Computer Vision and Pattern Recognition Workshops (CVPRW). pp. 1508--1516. IEEE (2022)

\bibitem{rw2019timm}
Wightman, R.: Pytorch image models. \url{https://github.com/rwightman/pytorch-image-models} (2019). \doi{10.5281/zenodo.4414861}

\bibitem{yu2023mix}
Yu, X., Wang, J., Zhao, Y., Gao, Y.: Mix-vit: Mixing attentive vision transformer for ultra-fine-grained visual categorization. Pattern Recognition  \textbf{135},  109131 (2023)

\bibitem{Zagoruyko2017}
Zagoruyko, S., Komodakis, N.: Wide residual networks (2017), \url{https://arxiv.org/abs/1605.07146}

\bibitem{zhou2019unet++}
Zhou, Z., Siddiquee, M.M.R., Tajbakhsh, N., Liang, J.: Unet++: Redesigning skip connections to exploit multiscale features in image segmentation. IEEE transactions on medical imaging  \textbf{39}(6),  1856--1867 (2019)

\end{thebibliography}
\end{document}